# Metatheoretical multiverse analysis: Improving the reliability of research with theories-as-data

**Nate Breznau**, breznau.nate@gmail.com, https://orcid.org/0000-0003-4983-3137

**Hung H.V. Nguyen**, https://orcid.org/0000-0001-9496-6217

German Institute for Adult Education – Leibniz Institute for Lifelong Learning
Heinemannstrasse 12-14
53175 Bonn, Germany

## Abstract

Here we present a method for measuring, analyzing and reducing theoretical uncertainty. We call it metatheoretical multiverse analysis (MMA). The method is important because the reliability and replicability of scientific observation and testing of a phenomenon are a function of theoretical uncertainty – thus reducing it is a method for improving theory. The MMA method requires encoding theory as propositional logic between variables (nodes) and their relationships with one another (edges). Propositional logic models are treated as causal path models in our method, which enable mathematical properties of testing and identification underlying their causal propositions. By encoding theories-as-data including their unknown components, it is possible to generate a multiverse of alternatively plausible theoretical models which can then be meta-analyzed using techniques borrowed from (statistical) multiverse analysis. We introduce metrics for studying theoretical multiverses that enable measuring and identifying the causes of metatheoretical uncertainty. This guides researchers to where it is best to invest theoretical development to reduce uncertainty. We use three simulations and a dedicated software package to demonstrate this method.

**Keywords**: Metatheory, Theories-as-data, Theoretical Uncertainty, Metatheoretical Multiverse Analysis (MMA)

**Statements and Declarations**: The authors declare no conflict of interest. Artificial Intelligence was employedwhat in research and writing. Initial ideas were co-brainstormed with a local agent using the BMAD Method. Multiple cloud and local agents acted as programming, data analysis, and writing assistants. We humans are responsible for the concepts, code quality, accuracy, integrity, editing, proofreading, and for all content, words and mistakes.

**Acknowledgements**: This method was developed with funding from the Germany Science Foundation (Deutsche Forschungsgemeinschaft) Grant Number 464546557. We are grateful for feedback obtained at the Institute 4 Replication's Barcelona Workshop, 2026.

## Introduction

Theory development is a primary goal of science. We want to know why we observe what we observe, and furthermore how to test if we know what we think we know (see discussion in Oppong 2022). It starts with a guess or conjecture and then through observation, testing and re-testing, a theory or set of theories develops. The reliability of theory is an inverse function of reproducibility. Low reproducibility equates to high uncertainty which means weak theory. High reproducibility on the other hand means strong theory, and perfect reproducibility means the theory is a law. There are many sources of uncertainty that can reduce reliability. Epistemic uncertainty results from varying ontologies and epistemologies, aleatoric uncertainty results from measurement error and sampling, and of course researchers make mistakes (Lakens et al. 2026). In this paper we address another form of uncertainty: theoretical uncertainty arising from the coexistence of multiple plausible alternative theories of the same phenomenon.

To investigate and reduce theoretical uncertainty we present metatheoretical multiverse analysis (MMA). This method encodes theories-as-data based on the underlying propositional logics in the theory. Crucial are the possibilities of multiple plausible theories and unknowns that exist in any given theory. From theories and their unknowns we generate a multiverse of plausible theories. We develop various measures to meta-compare the theories in a multiverse. Multiverse metadata and further analyses of the metadata provide methods for achieving myriad research goals. Our primary focus is discovery of the sources of theoretical uncertainty. Identifying these sources provides actionable targets for researchers' theory development efforts. The method thus economizes the theory building process. Researchers could use MMA for a variety of other goals like understanding theoretical uncertainty, developing better robustness checks and discovering new theories.

In this paper we introduce MMA and offer three simulations of how researchers might use it. These are a) *consensus illusion*, where many theories agree about core causal mechanisms but lurking in the structure of their multiverse are identification incompatibilities, 2) *uncertainty cruxes*, where analyzing a multiverse's components reveals how much weight they have in causing theoretical uncertainty, and 3) *ghost discovery*, where the multiverse is analyzed spatially using cluster analysis, and in one zone are extant mainstream theories, but hidden in other zones are other theories with completely different structures but which could constitute an alternatively plausible family of theories. Our simulations are in no way exhaustive of the fascinating potential ways of analyzing theories-as-data, and instead provide an introductory excursus. The MMA method has a dedicated R package running Python called *theoRy* used for producing the examples herein (Nguyen and Breznau 2026).

## Background

We define metatheoretical uncertainty as ambiguity caused by the simultaneous existence of multiple plausible theories that explain the same observed phenomenon. By phenomenon we mean anything: a process, event, trend, effect, etc. To keep our approach as parsimonious as possible we call a phenomenon $Y$ and our theories will explain why we observe $Y$, namely its causes which we label as '$X$' with numerical subsets. The MMA approach zeros in on an exposure, a focal causal variable $X_1$. Theories of the $X_1, Y$ relationship need to identify $X_1$ as a cause. To understand if a theory can correctly identify $X_1$ as a cause requires knowledge of other factors that cause $Y$ that might also link in some way to $X_1$. Most of our readers will be familiar with the idea of confounding for example, where a third variable causes both $X_1$ and $Y$ that could bias identification of causality if not included in the model.

For applied social and behavioral researchers, a quick example with the broken windows theory may be pedagogically useful. This theory postulates that the more disorderly a neighborhood ($X_1$), the more likely crime occurs ($Y$). As proposed by Kelling and Wilson (1982) it argues that dilapidated neighborhoods discourage social control which causes criminal activity. Following this theory is a competing theory suggesting that lead ($X_2$) is present in the buildings of more 'run down' neighborhoods and this causes developmental disorders that cause crime (Nevin 2000). Because $X_1$ and $X_2$ both refer to dilapidated neighborhoods, we cannot distinguish one theoretical explanation from the other if we only include dilapidated neighborhoods in our model. Moreover, other theories suggest drugs ($X_3$) (Blumstein and Wallman 2006), abortion ($X_4$) (Donohue and Levitt 2001) and demographic aging (neighborhood age segregation) cause crime ($X_5$) (Oberwittler and Svensson 2025) and these are known to frequent 'broken windows' neighborhoods.

If there are five equally plausible explanations for the outcome of crime ($Y$), then there is a 20% (⅕) chance that any one is correct, all else equal. If we add even more plausible theories it will push the likelihood toward zero asymptotically. With no other selection criteria, Occam's logic suggests 'the simplest explanation is the correct one'. But this is choice optimization and not necessarily useful in theory development. Parsimony is orthogonal to real-world accuracy and truthfulness. Therefore, MMA does not encourage taking the simplest theory as the best, and rather encourages interrogating all theories equally, regardless of how complex.

All phenomena in the social and behavioral sciences, like crime, are extremely complex and this means they tend to suffer from weak theoretical explanations (Scheel 2022; Willer and Emanuelson 2020; Meehl 1990). They have too many plausible alternative explanations and a paucity of studies to distinguish them from one another. Therefore, it is not surprising that there is a replicability and reliability crisis in social and behavioral research (Miske et al. 2026; Baker 2016). A theory that unlikely explains the phenomenon generates noise rather than confirm or disconfirm the theory. Without sound theory, we do not know the data generating model. Without this knowledge we are less likely to produce reliable and reproducible knowledge. It is an endogenous scientific trap that keeps many disciplines from moving forward.

The solution is to improve theory. However, scholars do not spend time improving theory necessarily where it is most needed. Instead they invest in what is 'hot', or what their supervisors and other gatekeepers want, thus creating institutionalized theory-pipelines that limit the scope of possibilities (Abbott 1999; Bourdieu 1975). Even if scholars could freely invest in theory, we do not know where their efforts are most needed. This means the development of theory is highly idiosyncratic and path dependent. It is likely that theory is desperately needed in certain areas but never gets developed. This is where MMA comes in.

## Theory

In theory building, it is crucial to investigate all possible $X$-variables and their associations with one another, to know if we are correctly identifying a causal relationship between $X_1$ and $Y$. Therefore, by "theory" here, we mean the explanation of $Y$ in sofar as this explanation can be used to design a study to effectively test for an effect of $X_1$ on $Y$ or not. In other words, whether we can safely conclude that we understand the causal $X_1 \rightarrow Y$ process or not, based on the other plausible theories that exist.

The MMA method is built on propositional logic, which is causal in nature because *if* $X_1$ *then* $Y$ means that the existence of $X_1$ guarantees $Y$. This is the most basic definition of causality, but it does not fit all forms of causal inference logic because there could be an unobserved cause of both $X_1$ and $Y$ that is causing the if-then relationship. In MMA there should not be unobserved confounders. Either these are known possibilities and they enter the multiverse as another variable like $X_2$, or they enter the multiverse as an unknown component that theoretically could exist. Thus, MMA is a method of both existing theory and unknown but plausible theory.

We aim to develop a method that is as agnostic as possible to the philosophical debates on causality that occur in the new and growing 'causal inference' subfield of science (Markus 2021). This is essentially impossible and we hope the user can set aside their priors about causality and causal inference methods to enable first steps toward algorithmic metatheory development. We settle on a working theorem that propositional logic statements like 'if $X_1$ then $Y$', or 'if $X_1$ then not $Y$', or 'if $X_1$ unknown $Y$' can be equivalent to arrows in causal path models. The basic graphical logic of causal path modelling, also present in directed acyclical graphs (DAGs), provides a way to encode theories-as-data and meta-analyze them.

The level of abstraction of a theory is $Y$-dependent. The $Y$ outcome can be a state, event, process or anything else that can exist or not exist or vary along some dimension of direction and intensity. It could be a person, institution or society for example. An outcome $Y$ might be crime rates, the ticking of "strongly agree" to a survey question, a decrease in wages, or an observed increase of voting for radical right-wing parties. If a researcher can imagine it, it is permissible as a $Y$ outcome.

## Theories-as-data

Theories come in various formats. The most common are semantic, descriptive and communicated via written or spoken language. In formal or causal modelling, a theory is instead expressed as an equation or graph. The difference is usually that descriptive text can account for nuance that is difficult or impossible to capture in a formula or graph (Abbott 2001). Consider intersectionality in the concept of 'the matrix of domination' (Collins 2000). The theory is that disadvantage operates via intersecting group boundaries both real and constructed (race, class, gender, skin color, religion, sexuality, etc.). The problem is that much of this process is subjective and unique to individuals. Intersectionality tends to reject categorization processes and seeks to replace this with intersubjectivity. This does not fit well with formulas or graphs which need concrete matrix elements and categories. We want to humbly make the reader aware of this in advance. In other words, before using our method the user has a potentially very difficult task of figuring out the propositional logic in their theories of interest.

We believe that all complex concepts can eventually be encoded if the user abstracts enough. It may be at a loss of accuracy, but think of the previous example: "Intersectionality" could be simply encoded as $X_1$. The user need not know how to measure it, only that it is theoretically possible to measure $X_1$. Then it becomes possible to use MMA, even if we do not know yet how to measure $X_1$. This is how concepts become variables ("nodes" in the language of causal path models) and the user can then specify what their theories say about the relationships of these variables to one another (causal arrows known as "edges" graphically).

The core of every theory in our method is the relationship of a cause or causal process, also known as exposure ($X_1$), with an outcome ($Y$). As with any form of causal logic, their relationship is temporal: $X_1$ occurs before $Y$. This rule introduces two theoretical time points into the encoding process. The basic components of an encoded theory are therefore nodes, edges and the timing of nodes. With these components, we have causal path models that reflect the propositional logic of theories and can combine them into a multiverse.

## A Theoretical Multiverse

With one or more theories encoded as data, we can construct a theoretical multiverse: the collective space of all causal specifications possible within a researcher's theoretical parameters. These parameters are nodes, edges and timings. With just one encoded theory model we can generate a theoretical multiverse if it has at least one additional $X$-variable (i.e., $X_2$) and we accept that we are not certain about the theoretical relationships of at least one or more nodes to another. In the broken windows example with social control ($X_1$) and lead ($X_2$), we might assume that a dilapidated neighborhood ($X_3$) causes both $X_1$ and $X_2$. Let us call this Model A. It would contain propositional 'if-then' claims as follows: $Model_A = \{X_1 \rightarrow Y, X_2 \rightarrow Y, X_3 \rightarrow X_1, X_3 \rightarrow X_2\}$.

The original broken windows theory does not include $X_3$ however. We call this Model B here: $Model_B = \left\{X_1 \rightarrow Y, X_2 \rightarrow Y, X_3 \overset{?}{\dashrightarrow} X_1, X_3 \overset{?}{\dashrightarrow} X_2\right\}$.The unknowns (?) associated with $X_3$ are an area where a researcher might want to invest time in developing theory. Figure 1 presents Model A and B as causal path models.

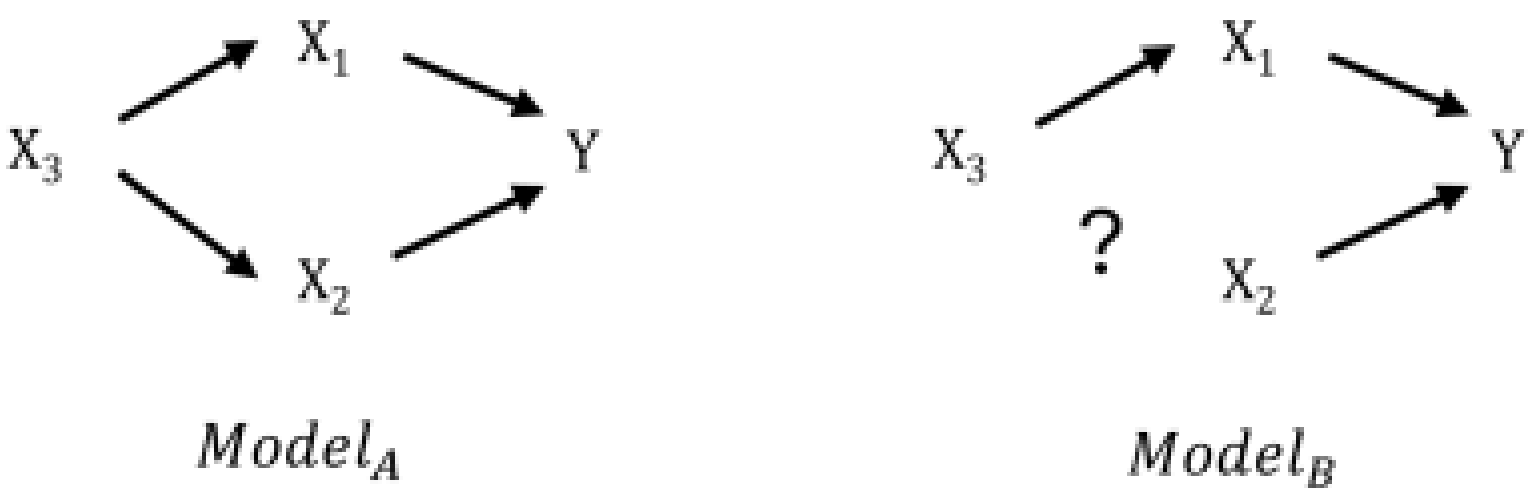

**Fig 1. Comparing different theoretical knowledge about $X_3$**

In Figure 1, it is easy to see that if $X_3$ causes $X_2$ as in $Model_A$ that it is a confounder. That means failure to include $X_3$ in a theoretical (or statistical) model takes away the causal identification of $X_1$ on $Y$. That is an argument to prioritize research on whether $X_3$ causes $X_2$ to move the theory construction process efficiently forward.

The method is flexible. The user could define further theories, for example that account for $X_4$ and $X_5$. They could re-specify this theory whereby $X_1$ occurs after $X_2$, or that a dilapidated neighborhood is not more likely to contain lead ($X_3$ does not cause $X_2$). Our method then can simulate all remaining unspecified relationships or counterfactuals, like flipping arrows to non-causal.

Introducing the universe of all potential possibilities where theory is unclear or competing theories do not agree, MMA maps out a metatheoretical space. Potential different theories have components with states of “absent” and “unknown” in our method. This creates the need for two types of node, *present* or *absent*, and three types of edges, *causal* (arrow, →), *non-causal* (no arrow in the case of a diagram, or slashed arrow $\nrightarrow$ in mathematical expressions) and *unknown* (question mark “?”, or question mark over dashed arrow $\overset{?}{\dashrightarrow}$, meaning not part of that theoretical model but of others in the multiverse like in Figure 1).

As economization of the theory building process is one of our primary targets, time invested in theoretical research into unknown edges that do not cause metatheoretical uncertainty would be wasted. The confounding problem in Figure 1 is an easy fact to deduce without a multiverse; however, as the number of nodes increases this becomes exponentially more difficult because so-called ‘back doors’ appear via highly complex pathways. Therefore, MMA is a method of comparing theories at scales not qualitatively possible to a human researcher.

Our pipeline for constructing a theoretical multiverse thus far: (1) establish a registry of all possible path model components, then (2) expand the registry into all possible theory model states, given timings or variation in potential timings. To analyze a multiverse then we recommend (3) dyadic compatibility between all theoretical models, and (4) mapping of zones in the multiverse.

**Component Registry**

The foundation of a multiverse are its nodes. We define the set of space-time nodes $N = \{(n_i, t_i) \mid i \in 1,2,\ldots,I\}$ where $i$ indexes the node order, $I$ denotes the total number of nodes, $t_i \in \mathbb{R}^+$ indicates the timing of node $n_i$. For a multiverse with only $Y$, $X_1$, and $X_2$ nodes, the largest possible $t_i$ is 3 because nodes can all occur at different times ($X_1$ causes $X_2$ causes $Y$), and the smallest is 2 ($X_1$, $X_2$ at the same time, $Y$ always after $X_1$). $t_i$ maximum always equals $I$. Note that it is possible for $X$ nodes (variables outside of exposure $X_1$) to appear after $Y$, making them potential colliders. The space-time node set is the basis of the node registry $R_{node}$ from which a registry of every edge ($R_{edge}$) between $n_i$ (the sending node) and $n_j$ (the receiving node) can be derived[1].

The *component registry* ($R$) is the combination of the node and edge registries which catalog the exhaustive collection of path model components that are theoretically valid within this multiverse. Edges that do not connect two nodes or nodes that have been excluded by the user are invalid. Formally:

$$R_{node} = \{(n_i, t_i) \mid i \in \{1,2,\dots,I\}\} \quad \text{Eq. 1}$$

$$R_{edge}^{\rightarrow} = \{(n_i, n_j) \mid (n_i, t_i), (n_j, t_j) \in N,\ i \neq j, t_i < t_j\}\ . \quad \text{Eq. 2}$$

Equation 2 means we only derive edge components (between a pair of $n_i$ and $n_j$) when temporal precedence allows. In other words, we require specifically that $\tau(n_i) < \tau(n_i)$, thereby preventing nodes from preceding their causes. In instances where timing is identical, bidirectional edge components (↔) are allowed. These indicate what in path modelling is known as 'residual covariance', and another way a user can specify that they are concerned about unobserved confounding but have no further information:

$$R_{edge}^{\leftrightarrow} = \{(n_i, n_j) \mid (n_i, t_i), (n_j, t_j) \in N,\ i \neq j, t_i = t_j\}\ . \quad \text{Eq. 3}$$

So that:

$$R_{(component\ registry)} = R_{node} \cup R_{edge}^{\rightarrow} \cup R_{edge}^{\leftrightarrow}. \quad \text{Eq. 4}$$

For our purposes, non-casual known paths are not included when listing the registry mathematically but they are none the less present.

The nature and utility of a multiverse depends on how it is 'seeded'. If a user puts in variables but does not specify further information, the multiverse will be very large and contain all possible models based on all possible positions and combinations of the concomitant nodes because every $t$ for node $n_i$ will include all possible timings $T$. This could be useful for simulations or exploring abstract metatheoretical space. But a focused use of MMA contains registry constraints. These could be defined simply by entering a single pre-existing 'seed' theory, or constraints that combine all theories in a subfield.

[1] Note also that temporal numbering does not have any meaning or anchor. It is just a system to generate lawful path models. In some models $Y$ might be at temporal position 1 and others at temporal position 6 in case there are five nodes all before $Y$ and none at the same time.

Registry constraints ($R_{constraints}$) are then the fixed parameters in the component registry, what is expected of a user serious about theory construction. For example, the 'broken windows' multiverse that would result from Figure 1 might have 18 models. It has three fixed edges, does not allow subsets of the four nodes or variation in their timing and includes the possibility of a bidirectional residual confounding of the nodes that occurred at the same time, see Table 1 and Figure A1.

We develop a shorthand so that we do not need to write out the entire registry for every multiverse. The nodes define all possible edges and if they are not listed in $R_{constraints}$ then they can take on any of the three states (causal, non-causal, unknown). Otherwise constraints on all states are listed as follows:

$$R_{node} = \{X_{4,t\in\{1,2\}}, X_{3,t=1}, X_{1,\mathrm{t}=2}, X_{2,\mathrm{t}=2}, Y_{t=3}\}^{2} \quad \text{Eq. 5}$$

$$R_{constraints} = \{X_4\{\rightarrow, \nrightarrow\}X_2, X_1 \rightarrow Y, X_2 \rightarrow Y, X_3 \rightarrow X_2, X_1 \overset{?}{\longleftrightarrow} X_2\} \quad \text{Eq. 6}$$

The set brackets designate things that are constrained to a limited subset. In Eq. 5, the multiverse will contain all versions of models with $X_4$ at time 1 and at time 2. In Eq. 6, the multiverse will contain all models with $X_4 \rightarrow X_2$ and $X_4 \nrightarrow X_2$ but not with $X_4 \overset{?}{\dashrightarrow} X_2$. In the case of double-headed arrows there is only present[3] or absent. A double-headed arrow holds the place of a theoretical confounder that is unobserved and is therefore by definition unknown.

## Model Registry

The *model registry* is every unique plausible path model that can be lawfully created from the component registry (meeting all criteria specified in Equations 1-4). All nodes are included in every model, even if they are not in every theory. When not in a theory they are absent for that model. A state of *absent* can only exist if that same node is *present* in at least one other model in the multiverse, otherwise the node simply does not exist at all in the multiverse.

---

[2] Reminder to the reader that the $X$ subset numbers are labels. They have nothing to do with the order or the nodes or their timings.

[3] Nodes occurring at the same time by definition cannot be causal so we use "present".

The registry is generated through exhaustive calculation; the Cartesian product of all possible models. Given $E$ components and their $|S|$ states and no fixed properties, the multiverse size is $M \leq |S|^E$. Again it makes little sense to simulate a multiverse where nodes can appear at every time point, this is the same as saying the researcher knows nothing about anything. Thus we expect multiverses to be smaller than $|S|^E$. We recommend allowing up to two nodes to vary in their timings and only in prespecified ways when constraining the component registry. For example, in the broken windows theory, it is possible that broken windows come after crime increases or before, and the user could specify this as a parameter in the multiverse generation process[4].

**Dyadic Compatibility Matrix and Theoretical Uncertainty**

Theoretical uncertainty arises when a phenomenon can be explained by alternative theories[5]. The data-generating model (correct theory, best 'truth') is not known as long as more than one plausible alternative theory exists. Theoretical uncertainty ($U$) is therefore our generic term to capture disagreement between models of theories-as-data. We define three non-exhaustive measures of $U$: $similarity_score$, $mas_compatible$ and $identified_compatible$. These are logical derivations and we suggest that researchers could and should measure $\Delta U$ in any logical way they find useful.

The basis for measuring theoretical uncertainty reduction $\Delta U$ is via comparing theories to one another. Hence the goal is to compare every theory-encoded-as-model to every other in the registry. We thus take all $M$ specifications in the model registry, and construct a dyadic matrix $D$ with all models as reference (ego, model $A$) and target (alter, model $B$). Every directed unique combination $(A, B)$, results in a total of $M \times (M - 1)$ discrete dyads.

*Similarity score*

---

[4] If the resulting multiverse too vast for computation, we recommend sampling. This can be random, or quasi-random Bayesian sampling that updates the space to fit posterior probabilities, to avoid missing 'zones'. This is a more advanced topic outside our current scope.

[5] Some might prefer "metatheoretical uncertainty", because "theoretical uncertainty" might be considered single-theory uncertainty, vagueness, porousness or structural issues. We argue these are also metatheoretical problems because they introduce alternative theoretical components (i.e., theories).

I might be useful to consider superficial structural similarity among theories. For this we propose a similarity score. It is defined by the number of identical components in the dyad, ($components_A \cap\ components_B$) meaning that the component (node/edge) is present in both models and identical in the case of the edge (same propositional assertion), divided by the total number of components in the dyad ($components_A \cup\ components_B$):

$$similarity_score(A, B) = \frac{components_A \cap components_B}{components_A \cup components_B}. \quad \text{Eq. 7}$$

The precise treatment of unknown edges and concomitant absent nodes in calculating this score depends on the dyad. In case of temporal variation of nodes, then $n_{i,t_i,model\ A}$ and $n_{j,t_j,model\ A}$ must be identical to $n_{i,t_i,model\ B}$ and $n_{j,t_j,model\ B}$ to be counted as 'identical'. When one model has the node or edge present, and thus not unknown, while the other has it as unknown, then the unknown could be repaired to create similarity. But if both models share an unknown edge, it is neither in the numerator nor denominator because one determines how the other must be changed to create similarity, and when one is unknown, the possibility for similarity remains unknown. The maximum similarity score is 1.0 for the case where all components are identical.

*MAS compatible*

If and how theory can identify causality ($X_1 \rightarrow Y$) is another property for comparison. In causal inference a Minimally Sufficient Adjustment Set (MAS) is a node combination required to correctly identify if the exposure caused the outcome (Pearl 2022). Larger models often have several MAS subsets, while some models have no MAS, meaning it is not possible to identify causality.

Therefore, if two models share at least one MAS, they have a special type of dyadic compatibility. We define this as $mas_compatible$ and let $MAS(m; X_1, Y)$ denote the collection of minimal adjustment sets returned for model $m$ for testing for a causal effect of $X_1$ on $Y$. Thus:

$$mas_compatible(A, B) = 1, MAS(A) \cap MAS(B) \neq \emptyset, 0, otherwise. \quad \text{Eq. 8}$$

If both have no valid MAS[6], or one has no MAS and the other does, they are not $mas_compatible$. At least one MAS must be shared because having no MAS means neither treatment effect is identified.

*Identified compatible*

Researchers working in the social and behavioral sciences often produce statistical models that throw in all variables as ‘controls’, sometimes known as the ‘kitchen sink’ approach. They likely have theoretical reasons that these ‘controls’ cause $Y$ in some way and so they include all of them, or they have secondary data that measured certain things and they use all these measures simply because they are available. Although these variables might cause $Y$ and seem theoretically relevant, the researcher does not specify the causal relationships between them and this could implicitly introduce identification problems. Our method is designed to interrogate this practice. Throwing all variables in might be ‘OK’, so long as they do not destroy causal identification of $X_1$ on $Y$. Therefore, when comparing two theory models we want to know first if they have the same nodes, and second if the entire model including all nodes and edges is identified with respect to $X_1 \rightarrow Y$.

Therefore, model A and B are $identified_compatible$ when:

1) Both models have the same nodes at the same timings.
2) Both models are identified: a model that includes (adjusts for) all its nodes, identifies a causal effect of $X_1$ on $Y$.

Compatibility means only that both theories taken at face value and adjusted as empirical tests would identify $X_1 \rightarrow Y$. It does not mean they will find the same parameter estimate, and crucially this metric does not consider what type of causal effect (total, direct, indirect). It is just a way to check if the theories’ full causal path models, without using any minimizing adjustment strategy of removing colliders or adjusting for confounders, can provide a successful causal test of $X_1 \rightarrow$

[6] There is an important distinction between no valid MAS and an *empty* MAS set. A simple model can return a valid empty MAS set. For example, where there is only exposure $X_1$ and outcome $Y$ and no other nodes, this is a valid, empty MAS. In cases where both models return valid, empty MAS sets, they are $mas_compatible$.

$Y$. Identification is evaluated separately for each model on its own graph, using standard causal-identification methods (Pearl 2010; Elwert 2013; Keele et al. 2020).

**Meta-Analysis**

With the component and model registries, and dyad matrix, we have a multiverse. Table 1 presents metadata from an example multiverse with four nodes, fixed temporal ordering, fixed edges reflecting Model A in Figure 1 and no node subsets allowed. This essentially captures the broken windows multiverse if it only had two theories in it – the original and the lead-causes-crime theory.

| Measure | Outcome |
| --- | --- |
| Component Registry | $R_{node}$ = $X_{3,t=1}, X_{1,\mathrm{t}=2}, X_{2,\mathrm{t}=2}, Y_{t=3}$<br>$R_{constraints}$ = $X_1 \rightarrow Y$, $X_2 \rightarrow Y, X_3 \rightarrow X_2, X_1 \overset{?}{\leftrightarrow} X_2$ |
| Models | 18 |
| Components | 10 |
| $mas_compatible$ | 9.8% (30/306 dyads) |
| Identified models | 18/18 |
| $identified_compatible$ | 100.0% (306/306 dyads) |
| Most common MAS | $X_2$ |
| MAS enabling most compatibility | $X_2$ |
| Uncertainty Cruxes<br>(Global Crux, top 3 reported) | 1. $X_1 \leftrightarrow X_2$, $\Delta U$ = 0.087<br>2. $X_3 \rightarrow X_1$, $\Delta U$ = 0.084<br>3. $X_3 \rightarrow Y$, $\Delta U$ = 0.084 |

**Table 1. Metadata for three node multiverse**

Note: Following the broken windows theory example where Y is crime, $X_1$ is broken windows, $X_2$ is lead and $X_3$ is dilapidated neighborhood. Furthermore, the bidirectional arrow reflects the possibility of other theories that confound this, where the brackets indicate not fixed but plausible. Crux $\Delta U$ measured here by $similarity_score$, but users could choose any metric in MMA. See Figure A1 for diagrams of all models. Produced using the theoRy package. Code available in repository https://github.com/hungnguyen167/theoRy/blob/main/vignettes/Table_1_AStA.R.

This multiverse contains 18 alternative plausible theory models which leads to 306 dyads. There are more than the provided two theoretical models because it remains unknown between them whether dilapidated neighborhoods $X_3$ really cause an increased incidence of lead, and unknown if there is unobserved confounding between broken windows and lead: $X_1 \leftrightarrow X_2$ (other than a dilapidated neighborhood). There are 10 components, 4 nodes and 6 edges including the allowance of a bidirectional edge $X_1 \leftrightarrow X_2$. Only 30 dyads contain MAS that are compatible. This MAS is simply $X_2$ because the role of $X_3$ can operate through $X_2$ and with the bidirectional edge, $X_2$ must remain in the model for identification. Otherwise, all models are identified if all nodes are included as an adjustment set (all confounding possibilities eliminated). The other features of Table 1 will be discussed shortly.

*Crux components*

Theoretical uncertainty exists because components differ between models. But the weight of these differences varies. Superficially using the $similarity_score$, components are all weighted equally if all models are present in the multiverse (all possible nodes, edges and timings); however, this should rarely be the case. Depending on the constraints applied, a component may appear more often in some models than in others. When it comes to identification using the $mas_compatible$ and $identified_compatible$ measures, certain components always matter much more than others, because they are key to identification. Components that cause the greatest $\Delta U$ we label *'crux' components* – those whose state change would have the largest impact on reducing theoretical uncertainty in the multiverse.

Our method addresses theoretical uncertainty at different levels of abstraction. Some theories' components are unknown. This arises because we simply do not know, or because some theories simply do not theorize about things that other theories theorize about. We propose two ways of thinking about and measuring the role of these unknowns. The first is about how much uncertainty having an unknown introduces. To address this we define *marginal crux* components. These exist only among components with a state of unknown. We can simulate how making this known would reduce dyadic incompatibility the most between two simulated states of causal ($S'_{causal}$) versus non-causal ($S'_{non-causal}$). This is computed by forcing edge $e$ to both possible known states for every model where it was previously *unknown*. Marginal cruxes are found by taking each unknown edge $e_{unknown}$, defined as an edge marked *unknown* in at least one model where applicable, and simulating it as causal and then non-causal. In both cases we recompute dyadic compatibility and take the state with the largest $\Delta U$[7], rejecting cases where $\Delta U$ is negative.

$$\Delta U(e_{unknown}) = max(S'_{causal} - S_0,\ S'_{non-causal} - S_0,\ 0) \quad \text{Eq. 9}$$

Marginal crux tells us the loss in theoretical certainty we incur by having unknown components in metatheory.

The second way of thinking about unknown is to consider that a component can be causal or non-causal, this is what happens when we have two different theories claiming two different causalities. We are interested to know what happens to our entire multiverse if this theoretical dispute gets resolved. This means we fix a component, regardless of its starting state, to a single state in all models in the multiverse. We call these *global crux* components that when fixed throughout the multiverse to a single state ($S'_{fixed}$), would reduce $\Delta U$ the most. Global cruxes are found by taking every edge regardless of its existing state $e_{global}$ and fixing it throughout the multiverse to a single state of causal or non-causal and then recomputing theoretical uncertainty and comparing it to the baseline multiverse.

$$\Delta U\left(e_{\text{global}} \mid t = n\right) = \max\left(S'_{\text{fixed},t=n} - S_{0,t=n},\ 0\right)\ ^{8} \quad \text{Eq. 10}$$

Thus, global crux tells us the gains to be had from knowing the true state of a single component.

Timing is treated differently in marginal and global cruxes. For a marginal crux, models are retained when $e_{unknown}$ has different node timings within the multiverse. That is because marginal crux is a theory-centric metric, which focuses on how a theory with an unknown influences $U$; whereas global crux drops all models from the multiverse that do not share the node timing associated with $e_{global}$ because it has become 'known' and should no longer have any alternative plausible states.

In Table 1, we report global cruxes and see that if we know with certainty that there is confounding component $X_1 \leftrightarrow X_2$ it would reduce 0.087 of uncertainty, in other words by increasing the $similarity_score$ by nearly 9% for all dyads. A researcher could define a global crux using any $U$ metric, for example via $mas_compatible$ and $identified_compatible$. We use $similarity_score$ here simply as a demonstration.

---

[7] Again, the user can select how they measure $\Delta U$, for example $similarity_score$, $mas_compatible$, or $identified_compatible$, or they can introduce a combination or their own metric.

[8] The floor at zero ensures that an edge whose resolution would introduce more uncertainty is not treated as actively harmful, but simply as offering no U reduction (very unusual case).

Like with multiverse statistical analysis (Young and Cumberworth 2025), it is likely that combinations of variables are the source of more explained variance than single variables. Therefore, we extend the single crux component logic to crux sets, where we calculate the joint impact of simultaneously resolving multiple components $k$ to discover possible synergistic reductions in $U$:

$$synergy_score(K) = \Delta U_{combined}(K) - \sum_{e \in K} \Delta U(e). \quad \text{Eq. 11}$$

The formula is essentially identical for marginal and global crux component seeking. A positive score means that resolving the marginal unknowns of fixing various edges across the multiverse for that components set produces a greater $\Delta U$ than resolving its individual components independently. This matters because individual $\Delta U$ rankings can obscure dependencies between theoretical claims. Two edges might each rank modestly on their own, yet together trigger a disproportionate shift in compatibility, for instance because they form a competing causal pathway whose alternative specification only becomes coherent once both are resolved. Conversely, a synergy score near zero indicates that the components contribute independently, with no interaction effect. By flagging synergistic sets, the framework highlights clusters of uncertainties that are worth investigating as a package rather than in isolation.

*Theoretical clusters*

We encourage thinking spatially about a multiverse. It is possible that different regions in the this space have different theoretical properties and by mapping this space a researcher could make theoretical gains. For example, most subfields have only a handful of theories. Many may look relatively superficially similar. However, the multiverse of alternative possibilities can be vast and contain entire zones of theoretical theories that do not exist in the extant literature. Potentially these zones might contain clusters of meaningful or somehow useful theoretical models. The point is that clustering algorithms can assist MMA by exploring the space in a theoretical multiverse. This is speculative. We imagine clustering based on the various metrics we introduced, or potentially by offering a reference theory to see how close models in the multiverse reflect this theory. Simulation 3 picks up this spatial logic.

## Simulating MMA in Practice

To illustrate economization of the research process, we apply an MMA pipeline to three distinct simulation scenarios. These examples showcase MMA's capacity to explore theoretical multiverses, identify sources of theoretical uncertainty and uncover new theoretical possibilities.

*Simulation 1: Consensus Illusion*

A 'consensus illusion' is a scenario where extant theories in a subfield look structurally similar and share a key identification property. They reflect a consensus about the data generating model at their core and disagree only about peripheral or minor components. Semantically they appear to describe roughly the same processes and mechanisms. When the propositional logic is closely interrogated however, there are lurking problems that might make them incompatible.

We first measure superficial consensus via the $similarity_score$ (Eq. 5). Again, this measures structural similarities, without measuring identification or causal process similarities. We simulate two multiverses of models with a high $similarity_score$ and perfect agreement on $identified_compatible$ (Simulation 1A) or perfect agreement on $mas_compatible$ (Simulation 1B). The point is to show that despite high similarity and identification agreement, the minimum adjustment sets might vary considerably (Simulation 1A) or that despite high structural similarity and identical MAS, the full models might face identification incompatibility (Simulation 1B).

Both Simulation component registries are in Table 2. In Simulation 1A, there are edges that we deliberately let vary to expose MAS incompatibilities. First, pre-exposure disagreement hinges on the edge $X_6 \rightarrow X_1$, which can vary between causal, non-causal, and unknown. If it is causal, then the model contains the backdoor path $X_1 \leftarrow X_6 \rightarrow Y$. This path confounds $X_1 \rightarrow Y$ and must be blocked by adjusting for $X_6$. If $X_6 \rightarrow X_1$ is non-causal, this backdoor does not exist, and $X_6$ is not in the MAS. In this simulation we exploit the fact that a single edge can produce disagreement in $X_1 \rightarrow Y$ identification, even when the theories agree on most or all other causal claims.

To create variation but keep similarity score high we allow six background edges to vary only between causal and non-causal. The first four determine whether $X_2$ through $X_5$ create additional backdoor paths into $X_1$. The final two vary the causes of $X_7$. Together, these six claims generate 64 different combinations. When combined with the three states of $X_6, X_1$dyad, the full multiverse contains 192 models. Within this constrained multiverse, we chose a reference theory where all seven relevant edges (the focal edge plus six background edges) are set to non-causal. The reference theory is used for clustering when presenting results (explained below).

Simulation 1B is set up to expose the consensus illusion stemming from a potential collider. The central disagreement is over the post-outcome edge $X_1 \rightarrow X_6$, which is allowed to vary between causal, non-causal, and unknown. Because $Y \rightarrow X_6$ is fixed as causal (see Table 2 for all constraints), when $X_1 \rightarrow X_6$ is causal, $X_6$ becomes a collider on the path $X_1 \rightarrow X_6 \leftarrow Y$. When conditioned on, this collider opens a path, effectively creating a statistical association between $X_1$ and $Y$ that would not otherwise exist – this is why the MAS can remain identical but the full adjustment sets are not. When $X_1 \rightarrow X_6$ is non-causal, the collider path does not exist, and conditioning on the remaining nodes blocks the confounding path through $X_2$. If $X_1 \rightarrow X_6$ is unknown, the theory has both a causal and non-causal completion; because identification does not hold in both cases, it also does not hold for the unknown case. Simulation 1B therefore shows how disagreement about a single post-outcome edge can produce disagreement about identification even when every model has the same MAS.

Similar to simulation 1A, there are also six background edges that are allowed to vary between causal and non-causal (see Table 2). Three describe relationships among $X_3$, $X_4$, and $X_5$, while the other three whether these variables cause $Y$. They introduce background structural variation across otherwise related theories. Together, these six binary claims generate 64 alternative theories. Combined with the focal claim $X_1 \rightarrow X_6$ (varying among three states), the constrained multiverse in simulation 1B also has 192 theories in total. In this multiverse, we choose a reference theory where all seven relevant edges ($X_1 \rightarrow X_6$ plus the six causal/non-causal only edges) are non-causal.

In both simulations, we report results across four cluster sizes around the reference theory. We define four cluster sizes according to the maximum number of background edges (out of six) that may differ from the reference model. The focused cluster includes the reference model and nearby models in which at most only one of the six background edges can deviate from the reference. This cluster contains 21 theories. The moderate cluster permits up to two edge deviations and contains 66 theories. The broad cluster permits up to three deviations and contains 126 theories. Finally, the full multiverse contains 192 theories.

**Simulation 1A**

Component registry:

$$R_{node} = \{X_{2,t=1}, X_{3,t=1}, X_{4,t=1}, X_{5,t=1}, X_{6,t=1}, X_{1,t=2}, X_{7,t=3}, X_{8,t=4}, Y_{t=5}\}$$

$$R_{constraints} = \begin{Bmatrix} X_1 \rightarrow X_7, X_7 \rightarrow X_8, X_8 \rightarrow Y, X_1 \rightarrow Y, X_2 \rightarrow Y, X_3 \rightarrow Y, X_4 \rightarrow Y, X_5 \rightarrow Y, X_6 \rightarrow Y, \\ X_2\{\rightarrow, \nrightarrow\}X_1, X_3\{\rightarrow, \nrightarrow\}X_1, X_4\{\rightarrow, \nrightarrow\}X_1, X_5\{\rightarrow, \nrightarrow\}X_1, X_2\{\rightarrow, \nrightarrow\}X_7, X_3\{\rightarrow, \nrightarrow\}X_7 \end{Bmatrix}$$

| Number of deviations allowed | Models | Mean similarity | MAS compatibility | Identified compatibility |
|---|---|---|---|---|
| 1 | 21 | 0.850760 | 0.028571 | 1.000000 |
| 2 | 66 | 0.794591 | 0.016783 | 1.000000 |
| 3 | 126 | 0.767506 | 0.012190 | 1.000000 |
| 6 | 192 | 0.757366 | 0.010471 | 1.000000 |

**Simulation 1B**

Component registry:

$$R_{node} = \{X_{2,t=1}, X_{1,t=2}, X_{3,t=3}, X_{4,t=4}, X_{5,t=5}, Y_{t=6}, X_{6,t=7}\}$$

$$R_{constraints} = \begin{Bmatrix} X_2 \rightarrow X_1,\ X_2 \rightarrow Y,\ X_1 \rightarrow Y,\ Y \rightarrow X_6, \\ X_3\{\rightarrow, \nrightarrow\}X_4, X_3\{\rightarrow, \nrightarrow\}X_5, X_4\{\rightarrow, \nrightarrow\}X_5, X_3\{\rightarrow, \nrightarrow\}Y, X_4\{\rightarrow, \nrightarrow\}Y, X_5\{\rightarrow, \nrightarrow\}Y \end{Bmatrix}$$

| Number of deviations allowed | Models | Mean similarity | MAS compatibility | Identified compatibility |
|---|---|---|---|---|
| 1 | 21 | 0.798270 | 1.000000 | 0.100000 |
| 2 | 66 | 0.725683 | 1.000000 | 0.107692 |
| 3 | 126 | 0.691166 | 1.000000 | 0.109333 |
| 6 | 192 | 0.678245 | 1.000000 | 0.109948 |

**Table 2: Multiverses presenting a consensus illusion of high superficial similarity and low theoretical compatibility.**

Note: Metatheoretical multiverse analysis (MMA) supported by the R package theoRy. Simulations present a high mean similarity theory multiverse using the dyadic $simularity_score$ (see Eq. 7) in four versions (‘neighborhoods’): Focused - only one background claim varies, Moderate - two background claims vary, Broad - three background claims, and Full - all theories for this simulated subfield included. Simulation 1A shows high structural similarity and identified compatibility but low MAS compatibility. Simulation 1B shows high structural similarity and MAS compatibility but low identified compatibility. They together represent the consensus illusion. A $\{\rightarrow, \nrightarrow\}$ symbol denotes that an edge is allowed to vary between causal and non-causal.

In Simulation 1A, mean component-level similarity remains between 75 and 85 percent at all levels of precision of the theory cluster, whereas MAS compatibility falls from 2.9 percent in the focused neighborhood to approximately 1 percent in the full multiverse. This low compatibility is a result of the strict definition of MAS compatibility: two theories are compatible only if they share at least one identical minimally sufficient adjustment set. Even a single disagreement about whether a pre-exposure variable causes $X_1$ can add or remove a variable from a given MAS. Partial theories (where nodes are absent) reduce compatibility further because the two resolutions of $X_6 \rightarrow X_1$ require different MAS. Meanwhile, identified compatibility remains perfect across all levels, since the multiverse setup does not contain any potential identification issues when all nodes are adjusted.

In Simulation 1B, mean component-level similarity remains between approximately 68 and 80 percent across all levels of precision, whereas identification compatibility hangs around 10 to 11 percent. MAS compatibility is 100 percent because every theory retains the same adjustment set $\{X_2\}$. Identified compatibility is much lower both theories in a pair must be identified under complete conditioning (adjusting for all nodes) and the lurking collider $X_6$ undermines this.

These two simulations demonstrate that despite high scores on metrics that might drive researchers to believe a family of theories are compatible, these theories can still contain major incompatibilities.

*Simulation 2: Crux Components*

We next consider a multiverse with great theoretical uncertainty because it has few knowns, i.e., few $R_{constraints}$. We look for crux components which could help with theory development, because their resolution would bring the greatest reduction of theoretical uncertainty ($\Delta U$). In Simulation 2 there are five variables: $R_{node}$ = $\{X_{2,\mathrm{t}=1}, X_{1,\mathrm{t}=2}, X_{4,t=3}, X_{3,t=4}, Y_{t=5}\}$. We know and constrain five causal edges: $R_{constraints}$ = $\{X_2 \rightarrow X_4, X_2 \rightarrow Y, X_1 \rightarrow X_3, X_1 \rightarrow Y, X_4 \rightarrow Y, X_3 \rightarrow Y\}$. This leaves four remaining unknown edges, which will take on unknown, causal and non-causal ($X_2 \overset{?}{\dashrightarrow} X_1$, $X_2 \overset{?}{\dashrightarrow} X_3$, $X_1 \overset{?}{\dashrightarrow} X_4$, $X_4 \overset{?}{\dashrightarrow} X_3$). Thus, the multiverse consists of $3^4 = 81$ models. For parsimony we chose models that are all $identified_compatible$ from the outset.

The unknowns become the source of testing for potential reductions in $U$ measured via $similarity_score$ and $mas_compatible$. Figure 2 shows component-specific returns for the four components that are allowed to vary within the multiverse. Resolving any of them would raise the $similarity_score$ by a negligible amount (0.50 percentage points, the purple bars). Each edge can be causal, non-causal, or unknown, and the multiverse includes every possible combination of those choices. Resolving any one edge therefore removes the same amount of disagreement between theories and increases structural similarity by the same amount.

Structural similarity measures whether theories agree about an edge, but not whether that edge matters for identification of $X_1 \rightarrow Y$.

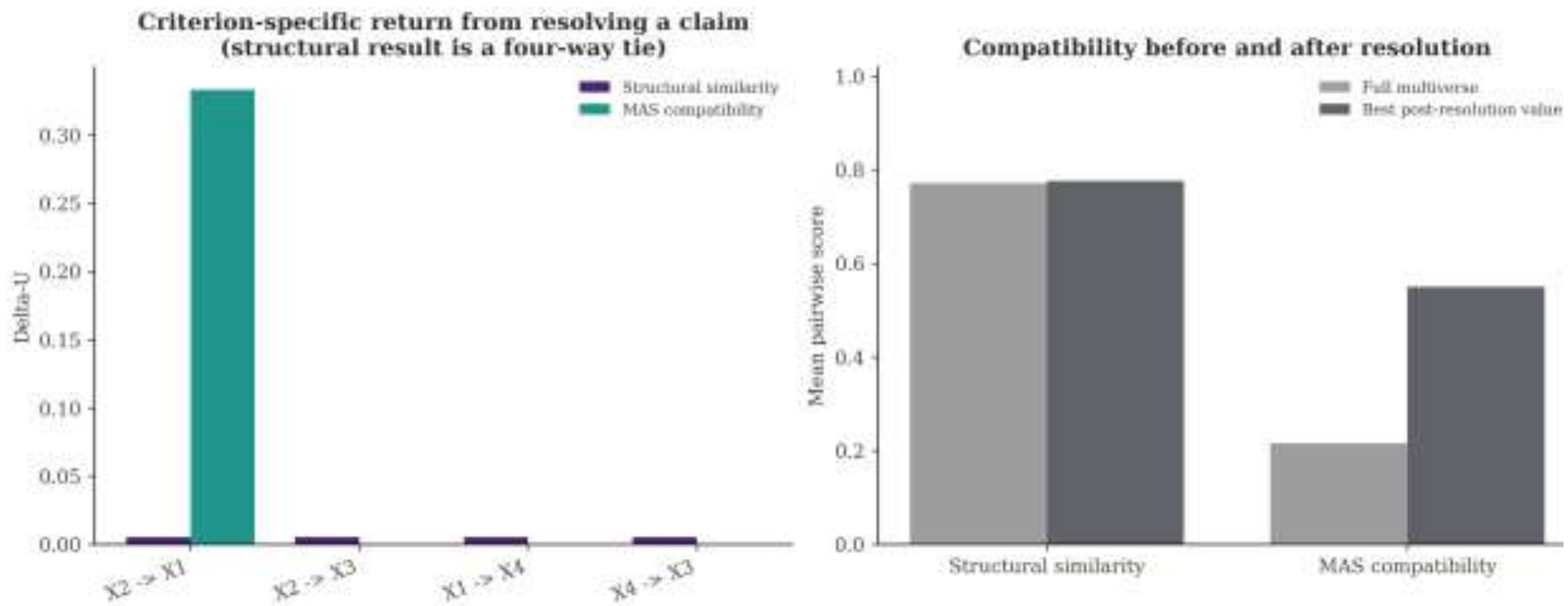

**Figure 2. Uncertainty reduction from an identified marginal crux component in a simulated multiverse**

Note: Panel A shows the reduction in theoretical uncertainty if one of four components becomes fixed (known) in the theory multiverse (the dark grey bars). Panel B shows the mean compatibility (1-U) score for $similarity_score$ ("structural similarity") and $mas_compatible$ ("MAS compatibility") if all four were to be known. The conclusion from Panel A is that $X_1 \rightarrow X_2$ is a "crux" component.

Where gains vary among unknown components is in $mas_compatible$. If $X_2 \rightarrow X_1$ is no longer unknown, it raises compatibility of Minimal Adjustment Sets from 0.22 to 0.55, a 33-percentage point increase. Because we searched for a marginal crux here (measuring $\Delta U$ as $mas_compatible$), this result is based on the edge no longer being unknown regardless of whether it is in the end causal or non-causal. This is an important piece of knowledge because theorists do not know in advance what they will discover when investing in theory, and this scenario considers both possible outcomes. Although a small multiverse, the results are powerfully suggestive of where a theorist could best invest further work. And the value of this increases exponentially with more complex multiverses where it becomes impossible to figure this out 'by hand'.

*Simulation 3: Ghost Discovery*

In our third simulation we imagine theories in a subfield that have structural and identification similarities, like in Simulation 1A. This time, however, the subfield theories have many more unknown components so that a multiverse simulating all possible states of those unknowns is massive. In our imagined subfield, the theories cluster in a family because they all share a high $similarity_score$ and reasonably high $mas_compatible$ and $identified_compatible$ scores. These are theories that have much in common in their identification of the $X_1 \rightarrow Y$ relationship. This theories of this family can be identified as near each other in multiverse space via a clustering algorithm. We call this the mainstream cluster, because it theoretically represents real, existing theories. However, the multiverse is so vast that the potential exists for other families of theories. These other families theoretically do not exist in the real subfield. They are an alternative family because they share low internal $U$ – high scores on all three measures – but high external $U$ – are not compatible with the mainstream family (or other potential families). If such a family exists, we call it a ghost cluster.

To demonstrate that ghost clusters are possible, Simulation 3 uses eight observed variables: $X_2$ and $X_3$ as baseline pre-exposure variables, $X_1$ as the exposure, $X_4$ through $X_7$ as post-exposure mechanisms, and $Y$ as the outcome: $R_{node}$ = $\{X_{2,\mathrm{t=1}}, X_{3,\mathrm{t=2}}, X_{1,\mathrm{t=3}}, X_{4,\mathrm{t=4}}, X_{5,\mathrm{t=5}}, X_{6,\mathrm{t=6}}, X_{7,\mathrm{t=7}}, Y_{\mathrm{t=8}}\}$.The registry includes every directed edge permitted by the temporal ordering, resulting in a total of 28 unknown edge components with three possible states each. The complete multiverse space is extremely large with $3^{28}$ theories. In real-world applications, a researcher would likely start with a full multiverse and then apply clustering with specific criteria fixed—such as specific theories from the literature. Here, instead we use a small sample of a multiverse to demonstrate how ghost cluster identification works, which avoids the need for a supercomputer and long computing times.

Our multiverse has a stratified 200-theory sample. The mainstream cluster is generated via a prototype. This is a single theory used for simulating other theories that are close to it structurally. Using the $R_{node}$ set from above, the prototype constrains a sequential pathway with direct paths from every non-outcome variable to the outcome:

$$R_{constraints,mainstream} = \left\{ \begin{array}{c} X_3 \rightarrow X_1, X_2 \rightarrow Y, X_3 \rightarrow Y, X_1 \rightarrow Y, \\ X_2 \rightarrow X_3, X_2 \nrightarrow X_1, \\ X_2 \nrightarrow X_4, X_2 \nrightarrow X_5, X_2 \nrightarrow X_6, X_2 \nrightarrow X_7, \\ X_3 \nrightarrow X_4, X_3 \nrightarrow X_5, X_3 \nrightarrow X_6, X_3 \nrightarrow X_7, \\ X_1 \rightarrow X_4, X_1 \nrightarrow X_5, X_1 \nrightarrow X_6, X_1 \nrightarrow X_7, \\ X_4 \rightarrow X_5, X_4 \nrightarrow X_6, X_4 \nrightarrow X_7, X_4 \rightarrow Y, \\ X_5 \rightarrow X_6, X_5 \nrightarrow X_7, X_5 \rightarrow Y, \\ X_6 \rightarrow X_7, X_6 \rightarrow Y, X_7 \rightarrow Y \end{array} \right\} \qquad \text{Eq. 12}$$

We generate the ghost cluster using an alternative prototype. Using the same nodes and timings, the ghost prototype has a very different causal structure:

$$R_{constraints,ghost} = \left\{ \begin{array}{c} X_3 \rightarrow X_1, X_2 \rightarrow Y, X_3 \rightarrow Y, X_1 \rightarrow Y, \\ \boldsymbol{X_2 \nrightarrow X_3, X_2 \rightarrow X_1,} \\ \boldsymbol{X_2 \rightarrow X_4, X_2 \rightarrow X_5, X_2 \rightarrow X_6, X_2 \rightarrow X_7,} \\ \boldsymbol{X_3 \rightarrow X_4, X_3 \rightarrow X_5, X_3 \rightarrow X_6, X_3 \rightarrow X_7,} \\ \boldsymbol{X_1 \nrightarrow X_4}, X_1 \nrightarrow X_5, X_1 \nrightarrow X_6, X_1 \nrightarrow X_7, \\ \boldsymbol{X_4 \nrightarrow X_5}, X_4 \nrightarrow X_6, X_4 \nrightarrow X_7, \boldsymbol{X_4 \nrightarrow Y,} \\ \boldsymbol{X_5 \nrightarrow X_6}, X_5 \nrightarrow X_7, \boldsymbol{X_5 \nrightarrow Y,} \\ \boldsymbol{X_6 \nrightarrow X_7, X_6 \nrightarrow Y, X_7 \nrightarrow Y} \end{array} \right\} \qquad \text{Eq. 13}$$

The two prototypes agree that four edges are causal: $X_2 \rightarrow Y$, $X_3 \rightarrow Y$, $X_3 \rightarrow X_1$, and $X_1 \rightarrow Y$. They also agree that six edges are non-causal: $X_1 \nrightarrow X_5$, $X_1 \nrightarrow X_6$, $X_1 \nrightarrow X_7$, $X_4 \nrightarrow X_6$, $X_4 \nrightarrow X_7$, and $X_5 \nrightarrow X_7$. They take opposite positions on the remaining 18 edges (in bold): nine are causal only in the mainstream prototype and nine are causal only in the ghost prototype. The two theoretical clusters imply different choices about which variables must be adjusted for when estimating the total effect of $X_1$ on $Y$. The mainstream requires controlling for $X_3$, whereas the ghost cluster requires controlling for both $X_2$ and $X_3$. The clusters therefore do not share the same MAS.

Similar to Simulation 1, we generate models that are both in the $3^{28}$ model multiverse and are randomly varying each of the 22 unknown edges with probability $p = 0.02$. If the resulting model duplicates one already in the subpopulation, it is discarded and another is drawn. This procedure produces unique theories that remain close to their family center (the prototype) while allowing limited variation within each family. It is possible but not guaranteed that the prototype remains in the subpopulation using this approach.

We sparsely represent the random remaining multiverse of theories on allowing each of their 28 edges to have causal and non-causal states equally likely, $p$ = 0.50.

With our 200-model multiverse we run a clustering algorithm using the 200 $similarity_scores$ that can be generated for each dyad with each model as the ego. This results in 200 sets of 200 $similarity_scores$ which are ordered (model-specific to the alter) including the 1.0 score for each ego with itself as alter in the dyad for completeness. We then use Uniform Manifold Approximation and Projection (UMAP) (McInnes et al. 2020), to create a two-dimensional map of these scores. This nonlinear dimension-reduction method leads models with close $similarity_score$ to appear near one another in two-dimensional space.

We then extract the clusters using DBSCAN, a density-based clustering method (Wang et al. 2019) that searches for areas containing several nearby points. It treats those dense areas as clusters and leaves isolated points unassigned. This is useful for ghost discovery because the number of theoretical families does not need to be specified in advance and resulting clusters must not include all data points.

DBSCAN allows us to use a cluster radius $\varepsilon$ as a cutoff. We specify $\varepsilon$ = 0.20,0.35 and 0.50. Comparing several radius values shows whether the ghost family remains visible when the definition of “nearby” in the multiverse space varies. The second criterion for a cluster next to its radius is the minimum number of models to qualify as a cluster which we set at four.

After retrieving clusters from DBSCAN, we then evaluate each cluster using three measures: average internal similarity, within-cluster MAS compatibility, and within-cluster identified compatibility. A cluster must score above 0.6 on all three measures to pass the common coherence requirement. Among the clusters that pass this requirement, a cluster is classified as mainstream when its average structural similarity to the reference theory is above 0.50. On the other hand, a cluster qualifies as a ghost when its average structural similarity to the reference theory is below 0.50 and its MAS compatibility with the mainstream group is below 0.50. A cluster formed by DBSCAN but failing the three-measure requirement is labelled fragmented. Models not in any cluster are labelled noise.

Figure 3 presents the results (see Table A1 in the Appendix for more information). In the left panel we see that, $\epsilon$ = 0.20, leaves many models unassigned to clusters. It finds the ghost cluster[9] which has an internal $similarity_score$ of 0.86 and mainstream similarity of 0.35. Within this ghost cluster, 10 are planted theories and 1 comes from the randomly generated models. At $\epsilon$ = 0.35, the number of noise models drops significantly to 12. The identified ghost cluster found all 12 planted ghost theories plus 2 heterogenous models. At the highest radius $\epsilon$ = 0.50, there are no more noise models, only three clusters (mainstream, ghost, and fragmented), and 15 models in the identified ghost cluster (12 planted plus 3 heterogenous).

In an empirical application, the true identity of ghost clusters is unknown. The fact that $\epsilon$ = 0.20 recovers most of the planted theories shows that the method is highly sensitive. In other words, it would be nearly impossible for a ghost cluster to exist in the way we define it, and go undetected in a multiverse[10].

[9] Only one cluster passes all three-measure requirement.

The right-panel of Figure 3 shows how clustering looks at $\epsilon$ = 0.35. While there are many clusters detected (9) and many have high internal structural similarity, only two (mainstream and ghost) passes the three-measure requirement and one qualifies as a ghost cluster. Some noise models are close to the mainstream and ghost clusters but are not assigned by DBSCAN.

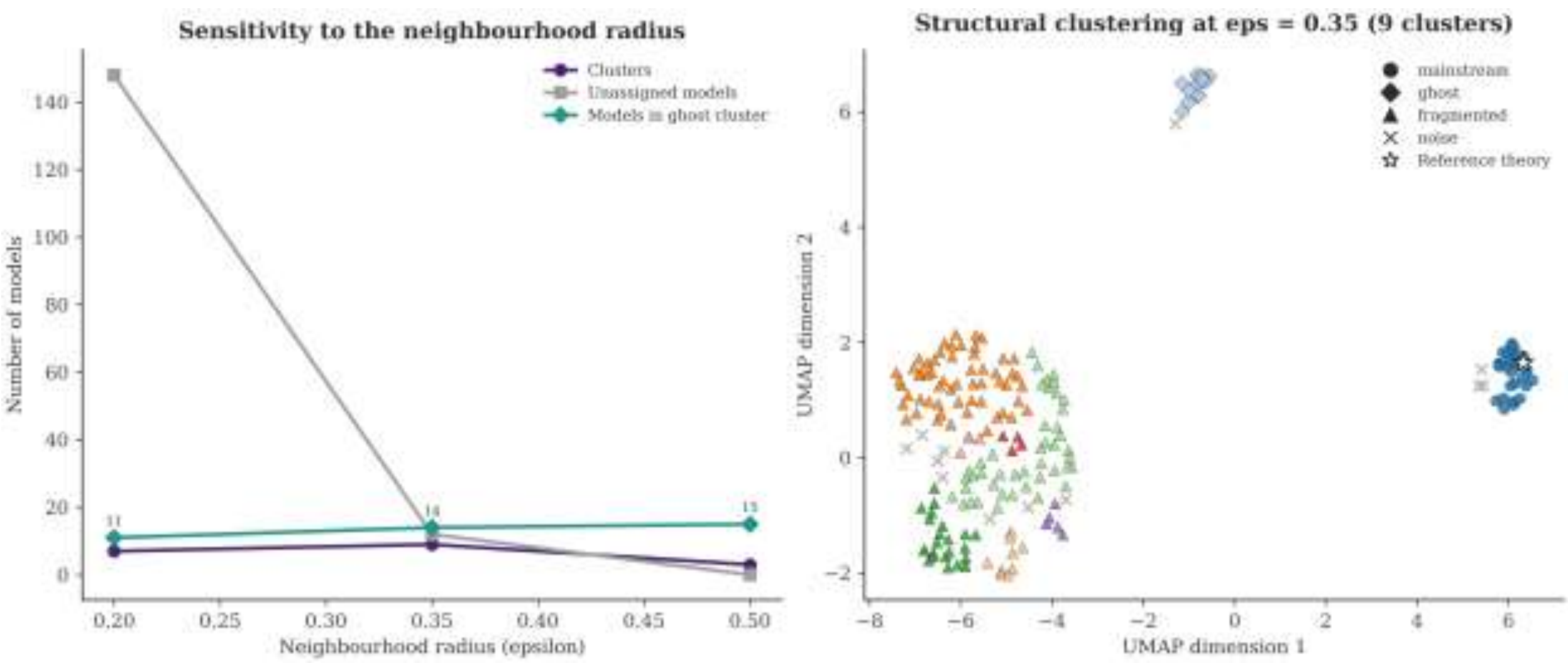

**Fig 3: Clustering theories in a multiverse with ghost cluster discovery**

Note: 200-model multiverse generated from two prototypes which seed the mainstream and ghost clusters, plus random generation to create a stratified sample of the $3^{28}$ model multiverse. Left-panel shows how many clusters are identified in the multiverse at three radius distances (purple), how many models are found in the ghost cluster (green) and how many models are not clustered ('noise'). Right-panel shows the two-dimensional UMAP distances using the sequential dyadic $similarity_score$ of all models. Fragmented are clusters that do not meet identification criteria (too much theoretical uncertainty) and noise are models that did not cluster. Code available at https://github.com/hungnguyen167/theoRy/blob/main/simulations/scripts/run_simulations.py

## Discussion

Metatheory Multiverse Analysis (MMA) enables comparison of theories encoded as data using their underlying propositional logics. Comparing theories has application in enabling and economizing the theory development process. Researchers can discover where to apply their efforts to get the greatest gains to theory; in other words, the greatest reduction in metatheoretical uncertainty. Researchers can also interrogate existing subfields to better understand their theoretical disagreements and possibly discover new theories. The MMA method has a dedicated R package *theoRy* (Nguyen and Breznau 2026) to facilitate the process.

---

[10] To further assess the reliability of the results, we generated 100 additional populations using seeds 1001 through 1100 and repeated the analysis with radius set to 0.50. A structural ghost was detected in every repetition.

Researchers need only convert semantic or formulaic theory into variables and their interlinkages encoded as nodes, edges and timings. This can be achieved in a number of ways and relies on the subjective knowledge and goals of the researcher – something researchers can find more details on elsewhere (Rohrer 2018; Ferguson et al. 2020). The user of our method need not commit to specific causal inference traditions. They only need to know if-then logic and in what order things happened. If they do not know the ordering they can also specify it as unknown and let the algorithm generate the resulting multiverse possibilities. Our method maps all theoretical uncertainties for the user giving them control over what they do not know, and how it may or may not threaten what they do know.

Another way of thinking about theoretical uncertainty is to imagine an applied empirical setting, with a researcher looking to identify a test of a causal effect of $X_1$ on $Y$. Perhaps a subdiscipline already has strong theoretical agreement, and this ostensible consensus means that using any of their theories to design a test will suffice. But maybe not. In Simulation 1 we showed how seemingly similar theories can contain lurking incompatibilities. Then it is important to look for ways to reduce this, otherwise scholars will not know what they are observing when they test if $X_1$ causes $Y$ 'in the wild'. One way to do this is to look for crux components, as shown in Simulation 2. A crux, is an economizing component for theory development.

Finally, as shown in Simulation 3, by identifying theoretical clusters that share strong similarities researchers can extend knowledge in a subfield or identify metatheoretical counterfactual theories. This could be valuable, because it leads to new discovery or uncertainty reduction. Theory could be weak because it mostly disagrees, or it could be weak because there is so little of it. MMA helps adjudicate this.

On the epistemological side, the development of the MMA method pushes a re-orientation toward metatheory. They way researchers methodologically engage their subjects will benefit from considering theoretical goals in advance (Scheel et al. 2021). Within the Open Science Movements a shift toward pre-registration, reducing questionable research practices and improving the robustness of estimates (Korbmacher et al. 2023) will further benefit from re-prioritizing larger theoretical goals. Theory should guide each stage of the research process, and if there is great theoretical uncertainty, this should be baked into the entire research workflow – possibly dealt with before launching into another expensive empirical endeavor.

Also, an epistemological promise of MMA is to highlight metatheoretical uncertainty as another form of uncertainty. This is something in addition to *aleatoric uncertainty* - the uncertainty associated with random sampling and statistical parameters of central tendency, and *epistemic uncertainty* - what is true in reality that cannot be altered through the observational and modelling activities of researchers (like the average height of a population, the number of reported crimes per capita, or the current percentage of the population that supports a candidate). The point is that as long as plausible alternative explanations exist for an observed outcome, there is uncertainty in how that outcome came to be that does not belong in the types of uncertainty we think about. This uncertainty cannot be removed without targeting it specifically through theory development (Laudan 1998). The reliability of empirical findings also depends on the reduction of this uncertainty, meaning MMA's theory development orientation can directly improve the credibility and reproducibility of research.

The application of MMA has limitations. The most glaring is the reduction of complex theoretical explanations into clean-cut logical models with clearly defined components. This is not a problem unique to metatheoretical work, this is a basic issue of the shift to causal inference that has taken place in recent decades. The distilling of complex, reciprocal, dynamic processes into linear, modular models is highly criticized (Cartwright 2007; Abbott 2001). The counter argument is that we would simply need more complex causal models to account for things like moderation, reciprocal processes and other complex semantic features. But even if scholars manage to encode theories successfully by abstracting, this process may leave them taking for granted many auxiliary assumptions about the theoretical process (Lakens et al. 2026). Without first tackling auxiliary assumptions, the background confounding or sources of uncertainty that could crash our identification assumptions, we are flying blind through the theoretical multiverse.

Currently the use of our method requires users to deploy the *theoRy* package in R or to program their own theoretical multiverse and analysis. Use of R is in no way a given in the social behavioral sciences. The package is also currently in its second iteration, meaning not all aspects of the method are easily deployed via a 'push-button' package. Moreover, the *theoRy* package can only handle a limited number of nodes (components) before the multiverse becomes unmanageable in size. Solutions that are geometric or Bayesian (Markov) in nature are still under construction. If not using R or our package, users would need to program their own MMA analysis from scratch, and this requires very high level computational social science skills that are rare in the social and behavioral sciences.

The user of our method will hopefully develop new ways of comparing theory once they begin constructing and thinking about theoretical multiverses. There is an incredible amount of space for improvement. We focused so far on causal identification, but researchers could further look at theoretical uncertainty by breaking down the types of causality a theoretical model identifies or is capable of identifying, for example total, direct and indirect effects. It would be possible for researchers to assign weights to theoretical components. Agnostic economizing of theoretical uncertainty is not the only factor that should drive theory development, for example pressing issues in society might demand theoretical development even if the specific components do not play a large role in causal identification, or even in the outcomes of applied research.

Our long term goal is to use MMA in combination with traditional multiverse analysis which is primarily a form of robustness checking and model dependency investigation (Steegen et al. 2016; Young and Cumberworth 2025). In the future we hope to map theoretical uncertainty onto quantitative uncertainty and use the two together to perform abductive research. If two theories are incompatible we might look at results from models testing these theories. If different models come to different quantitative results we have a strong motivation to resolve these theoretical perspectives. If not, we want to resolve them, but might get greater gains focusing our attention elsewhere.

**Contributor Roles Taxonomy (CRediT)**

Conceptualization: NB, HN
Data curation: HN, NB
Formal analysis HN, NB
Funding acquisition: NB
Investigation: NB, HN
Methodology: NB, HN
Project administration: NB
Resources: NA
Software: HN, NB
Supervision: NB
Validation: HN, NB
Writing – original draft: NB, HN
Writing – review & editing: NB, HN

## Appendix A. Glossary

**Aleatoric uncertainty:** The form of statistical uncertainty associated with random sampling, measurement error, and statistical parameters of central tendency.

**Bidirectional arrow:** A double-headed arrow (↔) representing residual covariance or potential unobserved confounding between two variables. These are only derived when the temporal positions of the two nodes are identical, meaning they occur at the same point in time (Eq. 3, Eq. 6).

**Causal arrow:** A directed edge (→) representing a known, asserted causal relationship where the source node temporally precedes and causally influences the target node (e.g., Eq. 2, Eq. 6, etc. throughout).

**Cluster:** Any form of grouping models based on their features around a real or mathematically defined centroid. In our specific terminology there are mainstream (extant), ghost (theoretical) and fragmented (not internally, externally or coherent enough to be useful) clusters (Table A1, Figure 3).

**Component:** An individual node (including the space-time of the node) or derived edge within the component registry (e.g., Eq. 4).

**Component registry:** The complete, mathematically bounded catalog of all theoretically valid node and edge components within a multiverse. It is formally defined as the union of the node and temporally admissible edge registries: $R = R_node \cup R_edge$ (Eq. 1-4).

**Crux component:** A specific theoretical component (or set of components) whose resolution yields the greatest reduction in overall theoretical uncertainty ($\Delta U$) across the multiverse (Eq. 9-10).

**Edge:** The graphical representation of a relationship between two nodes in a path model, representing propositional assertions that can take on causal (→), non-causal (↛), or unknown ($\overset{?}{\dashrightarrow}$) states (e.g., Eq. 2-4).

**Encoded theory:** ('Theories-as-data') A qualitative or semantic theory that has been formally abstracted and converted into propositional logic. Specifically, a structured set of variables, propositional assertions, and temporal timings, allowing it to be treated as a causal path model.

**Epistemic uncertainty:** The form of ontological uncertainty concerning what is true in physical reality that cannot be altered or resolved through the observational and modeling activities of researchers. Could guide theory construction, because epistemology guides the way semantic, propositional and causal logic are defined for a researcher.

**Focal exposure ($X_1$):** The primary causal variable under study whose total or direct effect on the outcome ($Y$) the researcher is attempting to test and isolate. In MMA so far, we always specify this as $X_1$ but researchers could theoretically change this to fit their needs.

**Fragmented cluster:** A cluster of models identified in a theoretical multiverse during spatial clustering due to proximity, but which fails to meet the strict similarity and MAS compatibility thresholds required to qualify as a ghost cluster.

**Ghost cluster:** An internally consistent family of models discovered through density-based clustering of the theoretical multiverse that has high internal compatibility, but is not compatible with mainstream theories in the subfield, representing a logically plausible but currently untheorized family of explanations.

**Global crux:** An edge component (*e_global*) that, when fixed to a single state across all models in the multiverse, yields the largest reduction in global theoretical uncertainty (Eq. 10).

**Identified compatible:** A state of compatibility where two models are both structurally and causally capable of identifying the total causal effect of $X_1$ on $Y$. Specifically, both models must declare the exact same set of remaining nodes after intermediate nodes on the directed path from $X_1$ to $Y$ are removed, and both must successfully identify the causal effect when adjusting for all their included nodes (Eq. 8).

**Mainstream cluster:**

**Marginal crux:** An unspecified, unknown edge component (*e_unknown*) that, when resolved to either causal or non-causal across all models where it was previously unknown, reduces dyadic incompatibility the most. It identifies the most critical immediate theoretical blind spot. (Eq. 9)

**MAS:** (Minimally Sufficient Adjustment Set) A minimal ccombination of covariate nodes required to block all backdoor paths and correctly identify the causal effect of the focal exposure ($X_1$) on the outcome ($Y$). Larger models can have multiple valid MAS subsets, whereas some models with colliders or unblocked confounders may have none (Eq. 8).

**MAS compatible:** A measure of dyadic compatibility where two models (*A* and *B*) are compatible if they share at least one identical Minimally Sufficient Adjustment Set (MAS) required to identify the causal effect of $X_1$ on $Y$. It is represented as a binary indicator: *mas_compatible(A,B)* = 1 if $MAS_A \cap MAS_B$ is non-empty, and 0 otherwise.

**Metatheory:** An analytical orientation where theories themselves serve as the primary unit of analysis. It involves meta-analyzing a field's theories as data to investigate, understand and identify theoretical uncertainty.

**Model registry:** The exhaustive set of all unique, plausible, and lawful path models ($M$) that can be constructed through the Cartesian product of all possible component states in the component registry, formally bounded by $M \leq S^E$ (Eq. 4).

**Node:** A graphical and mathematical representation of a theoretical concept or variable ($n_i \in N$) in a path model, where each node is assigned a specific chronological point to establish temporal precedence (e.g., Eq .1, Eq. 4-5).

**Path model:** A graphical causal model (such as a Directed Acyclic Graph or ADMG) that encodes the variables (nodes) and propositional linkages (edges) of a theory to evaluate causal identification (Eq. 4, Figure 1, Figure A1).

**Propositional logic:** The underlying logical system of MMA where causal claims are represented as conditional statements (such as "if $X_1$ then $Y$"), which are shown to be mathematically equivalent to arrows in causal path models (Eq. 2).

**Registry constraints:** The fixed parameters (such as designated node timings or restricted edge states) applied to the component registry to limit the theoretical multiverse's expansion and focus analysis on a specific subfield's consensus (Eq. 5-6, Eq. 12-13).

**Similarity score:** A structural, superficial metric used to compare the similarity of two models (*A* and *B*) based on the ratio of shared identical nodes and edges to the total components in the dyad (Eq. 7).

**Synergy score:** A metric used to discover joint, non-independent effects of simultaneously resolving a set of components ($K$) to identify compound theoretical uncertainties that are worth investigating as a package.

**Timing:** Each node in the set $N$ must have a specific temporal point (time $t$) linked to it, but can have multiple times. Nodes cannot exist without timing in a multiverse (Eq. 1-3).

**Theoretical multiverse:** The collective, multidimensional space of all possible causal model specifications (encoded theories, theories-as-data) that can be lawfully generated based on a researcher's specified parameters of nodes, edges, and timings.

**Theoretical uncertainty:** The systematic ambiguity and disagreement that arises when multiple plausible alternative theories co-exist to explain the exact same observed phenomenon ($Y$). Also referred to as metatheoretical uncertainty. Designated as $U$, and change (reduction) in uncertainty ($\Delta U$) (Eq. 9-10).

**Theories-as-data:** The foundational paradigm of the Metatheoretical Multiverse Analysis (MMA) method where qualitative, verbal, or semantic theories are systematically translated into structured quantitative data. They are encoded theories as networks of variables (nodes) and their relationships (edges) using propositional logic (Eq. 4).

**Theory:** The substantive explanation of an outcome ($Y$) insofar as this explanation can be used by a researcher to design an empirical study to effectively test for a causal effect of a focal exposure $X_1$ on $Y$.

**Variable:** The operationalized, measurable counterpart of an abstract theoretical concept, which is represented graphically as a "node" in a causal path model (Eq. 1, Eq. 5, etc).

## Appendix B. Supplementary Tables and Figures

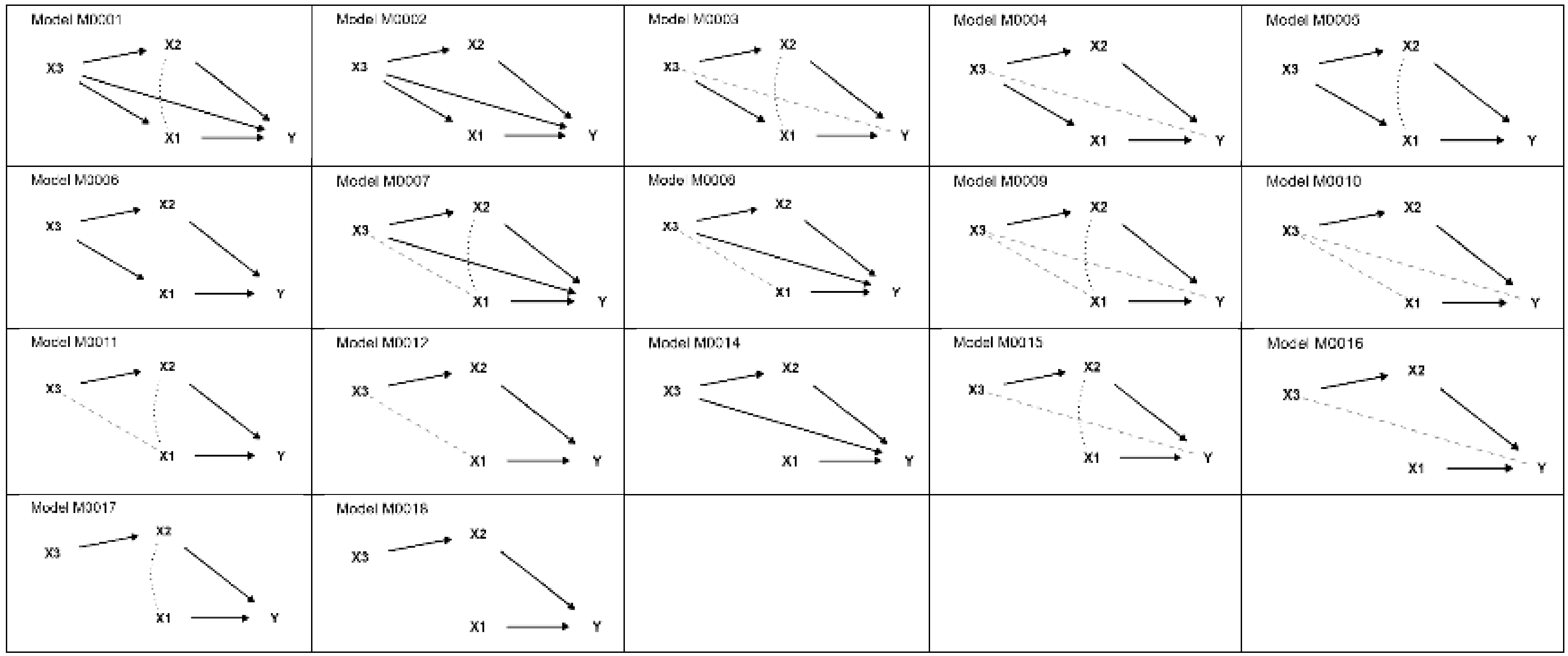

**Figure A1. Plots for Table 1 'Broken Glass' Multiverse**

Note: Causal path diagrams for all models in the multiverse summarized in Table 1. Arrows are causal (if-then proposition), lack of arrow is non-causal (if-not-then), grey dotted arrows are unknown and dotted line is a residual covariance (potential unobserved confounder). Multiverse generated via MMA method with fixed components $X_1 \rightarrow Y, X_2 \rightarrow Y, X_3 \rightarrow X_2$ using R package *theoRy*. Code available at https://github.com/hungnguyen167/theoRy/vignettes/Table_1_AStA.R).

| Radius | Clusters | Unassigned | Ghost-labelled models | Internal similarity (ghost) | Similarity to reference (ghost) | Within MAS (ghost) | Ghost-to-mainstream MAS |
|---|---|---|---|---|---|---|---|
| 0.20 | 7 | 148 | 11 | 0.860668 | 0.354532 | 0.818182 | 0.000000 |
| 0.35 | 9 | 12 | 14 | 0.835198 | 0.368973 | 0.857143 | 0.000000 |
| 0.50 | 3 | 0 | 15 | 0.815022 | 0.379836 | 0.742857 | 0.066667 |

**Table A1. Simulation 3 Results**

Note: Clustering is done with DBSCAN on UMAP two-dimensional coordinates at three radii: 0.20, 0.35, and 0.50. To qualify as a ghost cluster, the cluster must pass the three-measure requirement (structural similarity, MAS compatibility, and identified compatibility all above 0.60), with low similarity to reference model and low ghost-to-mainstream MAS (both under 0.50).